# A Compact High Frequency Voltage Amplifier for Superconductor-Semiconductor Logic Interface

Sasan Razmkhah[1], A. Bozbey[1], Pascal Febvre[2]

[1] Department of Electrical and Electronics Engineering, TOBB University of Economics and Technology, 06560, Ankara, Turkey.
[2] IMEP-LAHC, CNRS UMR5130, Université Savoie Mont Blanc. Le Bourget du Lac, France

E-mail: sasan.razmkhah@gmail.com

**Abstract**

The many advantages of cryogenically-cooled Single-Flux Quantum (SFQ) circuits imply that the high speed and low voltage output signals must be amplified and interfaced with standard electronics. State-of-the-art low-noise and wide-band amplifiers are required to interface with room temperature electronics. One solution is to place preamplifiers at the cryogenic stage, which requires specific semiconductor design and fabrication. However, a more viable and energy-efficient approach is to integrate the pulsed logic circuit output stages with on-chip superconductor preamplifiers. We designed, fabricated, and tested an on-chip compact voltage multiplier integrated with the output stage of SFQ circuits to increase the voltage amplitude of SFQ pulses. The circuit is designed with the same technology as the logic circuit hence its noise level is lower, and it works at higher frequencies compared to CMOS amplifiers and due to quantized nature of it there is no added noise. The fabricated circuit has a compact size of 160 µm×320µm and provides about 10 dB gain with measured 600 µV output voltage at frequencies up to ~25 GHz in simulations. By stacking more levels, over 20 dB gain at circuit level is achievable as shown in simulations. Moreover the gain of the superconductor voltage amplifier is quantized and programmable.

## 1. Introduction

SFQ-based superconductor circuits feature low noise, very low energy consumption and higher speed [1–3] compared to other technologies. However, the low voltage level of the output pulses of SFQ circuits at cryogenic temperatures, of the order of a few hundreds of microvolts, requires amplification to be further processed with room-temperature Complementary Metal-Oxide Semi-conductor (CMOS) electronics [4]. One of the reason is the lack of fast and compact memory in SFQ-based circuits [5,6], which requires for the moment CMOS memory to store and recall data coming from SFQ circuits.

Nevertheless, room temperature CMOS electronics requires a higher voltage difference between levels of at least 150mV to read the outputs of SFQ circuits. The output voltage of SFQ circuits is determined by the Josephson junction critical current and normal resistance. For the standard STP2 process fabrication technology of the AIST CRAVITY Foundry used in this work, $I_cR_N$ is 380 µV [7]. This is related to the pulse voltage level whose average DC value is around 200µV. This means that we need at least 60dB of voltage gain to reach a voltage level difference of 200 mV with differential output that CMOS circuits can analyze in presence of noise.

The outputs can be amplified at room temperature after the cryocooler stage but at the cost of additional noise since the noise of the wirings is also amplified and can be overwhelming at high sample rates, especially because the wiring length can be of several tens of centimeters which is the distance between the cryogenic circuit and room-temperature electronics. Besides the added noise at room-temperature is always higher than the one-off cryogenically cooled amplifiers.

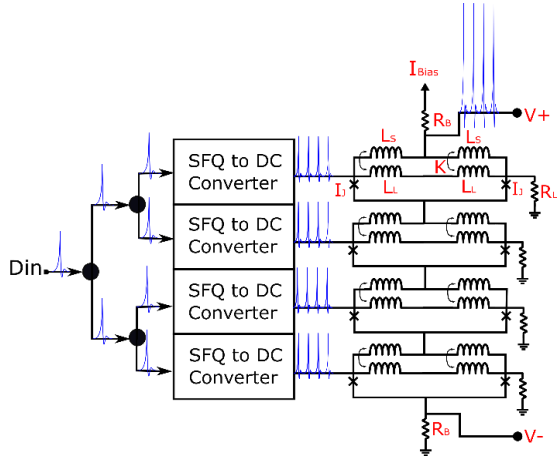

Figure 1 Schematic of the voltage multiplier circuit. The output of the multiplier is a differential voltage generated by the series of SQUIDs. Splitters and DC/SFQ converter cells are standard CONNECT library cells. The values for the SQUID stack are: $L_S$=4.7pH, $L_L$=4.7pH, K=0.5, $I_{Bias}$=230μA, $R_L$=2.1Ω, and critical current of the junctions $I_J$=120μA.

Many CMOS amplifiers for cryogenic temperatures, with different technologies were proposed for this purpose [8–13]. The most recent works are focused on the Si-Ge technology since it shows good response at cryogenic temperatures [14,12]. The last proposed design uses an amplifier on the 4K stage directly on the chip package [15]. However, there are several problems with such a high gain value at these unconventional temperatures. First, for a ~60dB gain the amount of bandwidth is strongly reduced (the gain × bandwidth product of an amplifier is fixed for a given technology). This reduces by a large amount the maximum speed of SFQ signals as well, which is one of the main appealing features of SFQ technology. Another problem is the lack of correct models for CMOS technology at ultra-low temperatures for high frequency applications [16,17]. This makes the design very difficult and costly. And the last problem is the price: an amplifier with 60dB gain at around 10GHz of bandwidth is required for each output channel. This leads to a high cost for circuits with tens of outputs such as arithmetic logic units or decoders.

One way to tackle this problem is to use on-chip cryogenic pre-amplifiers. By making a compact quantum accurate pre-amplifier based on superconductor technology that delivers output signals higher than 2 mV with a bandwidth of about 10 GHz, it becomes possible to use room-temperature CMOS amplifiers [18] with higher threshold levels. Therefore the power consumption drops significantly at the cryogenic stage. Several efforts have been made to develop on-chip amplifiers. Cryotrons and nano-cryotrons can be utilized to increase the output current of SFQ circuits[19]. While nano-cryotrons are promising as a cryogenic replacement for Field-Effect transistors (FETs) to amplify the current, the lack of switching speed and the difficulties of reproducibility in the fabrication process limit their practical use. Another cryogenic and on-chip method is obtained by stacking Josephson junctions [20–22] or SQUIDs [23–25]. Suzuki et al. [20] first presented the concept of high voltage readout for SFQ circuits by connecting their outputs to a series of shunted Josephson junctions that could drive a MOSFET switch. Smaller junctions drive larger ones and gradually the output voltage of the junctions increases. The limitation of this method is the dependence of the output speed to the switching speed of the MOSFET and the CMOS buffer stage. There was also the need for a hybrid superconductor-semiconductor fabrication process on the chip. Ortlepp et al. [22] have improved this structure in their work and removed the need for the on-chip MOSFET switch. This design has also its own limitations and while it is compact, the output voltage is limited. Kornev et al. [24] proposed Superconducting Quantum Interference Filters (SQIFs) to amplify pulses. In this method, the flux storage capacities ($L \times I_C$) of each SQUID loop of SQIFs are prime multipliers of the main SQIF loop capacity. While this method can be fast and needs low bias, it is not easily scalable since each SQIF needs to be designed from scratch. Herr et al. [23] uses a structure of SQUIDs to accumulate fluxons and release them at the same time to get a large pulse amplification. This method is also complicated to implement and scale and is not fast enough for interfacing purposes.

In the work presented by Soloviev et al. [25] the authors use a series of SQUIDs for impedance matching and to increase the voltage level. The use of SQUIDs circuits where each SQUID can be individually switched solves some of the limitations of the former work, as shown in [23–25]. In [25] the SQUIDs are connected to JTLs whose output pulse trains are added to achieve amplification. In this method, speed is limited: 1Gbit/s for output voltages of about 2.4 mV (20 dB gain). Also the sizes of SQUID loops that match JTLs are larger.

In this work we propose a new approach to increase the output gain by incorporating an SFQ-to-DC converter cell [26] whose intrinsic SQUID loop is utilized to control a larger coupled SQUID that performs the voltage amplification. Different circuits have been designed by placing in series several coupled SQUIDs to increase the output voltage. Some of them were fabricated and measured.

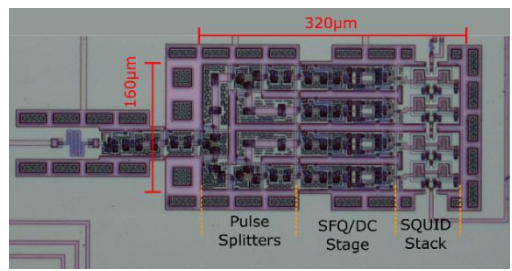

Figure 2 Layout of the fabricated circuit of the voltage multiplier using STP2 process [7]. The standard cells that are used are inspired from the CONNECT cell library **[27]**.

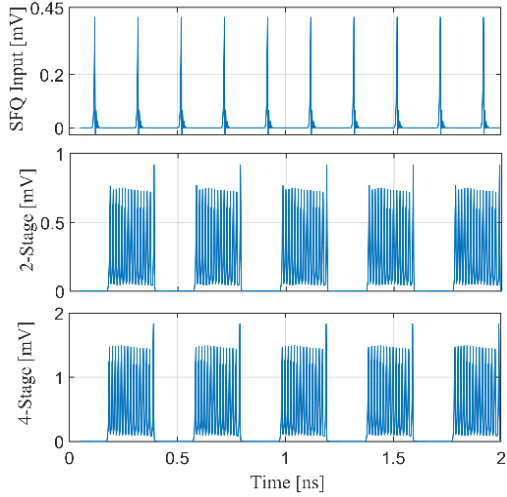

Figure 3 Simulation result of the input pulse and output voltage of voltage multipliers with 2 and 4 stages.

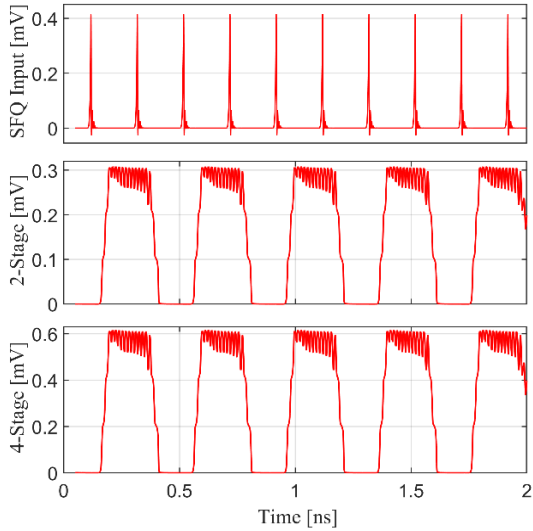

Figure 4 Simulation result of the input pulse and averaged output voltage (by low pass filtering) of multipliers with 2 and 4 amplification stages.

## 2. Methodology and Design

SFQ pulses carry a fixed amount of magnetic flux known as the magnetic quantum of flux, $\Phi_0$. Therefore, to amplify the voltage output of a pulse, we should either decrease the pulse duration, or increase the number of simultaneous output pulses per each input pulse and arrange them to add up through a dedicated circuit. In practice though, the pulse duration is fixed by the technology and the $I_cR_N$ product. Hence there is no easy way to increase substantially the output voltage this way. The second method to amplify the output voltage consists of increasing the energy of the output signal by stacking pulses on each other. In this case the ultimate output voltage is limited by the number of pulses that can be stacked. To do so, each input pulse must generate multiple pulses with the same timing and the voltages of these pulses should be added to produce the amplified output voltage. Figure 1 shows the schematic of the proposed and designed circuit. The first stage is a series of splitters that generates multiple pulses from a single SFQ pulse, with the same delay. The second stage is an SFQ-to-DC converter circuit which contains a SQUID loop at the output of its circuit [27]. The dc current of this loop toggles for each incoming pulse between a bias value lower than the SQUID critical current (associated to zero voltage) and a current value higher than the SQUID critical current, corresponding to an oscillating state of the SQUID. This oscillation is associated to a train of voltage pulses whose average DC value is larger than zero. Consequently, the SQUID voltage starts or stops oscillating as the state of the SFQ-to-DC converter changes for each incoming pulse, to produce a Non-Return-to-Zero (NRZ) voltage compatible with CMOS electronics standards.

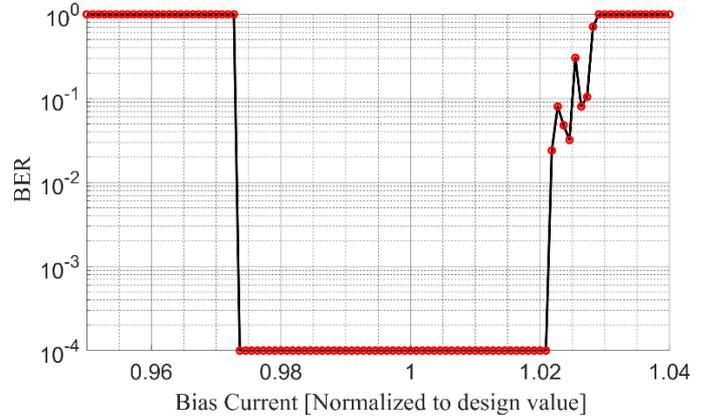

Figure 5 Bit error rate simulation of the 4-stage voltage multiplier versus normalized DC-SQUIDs bias current.

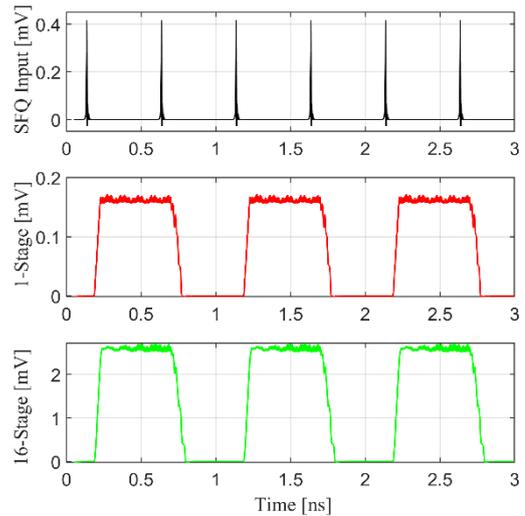

Figure 6 Simulation result of the input pulse, and averaged output voltages (by low pass filtering) of multipliers with 1, and 16 stages.

The last stage of the circuit is a series of SQUID loops. The SFQ-to-DC output is galvanicaly connected to the ground with a series RL circuit. The resistor is there to cut the superconductor loop and avoid accumulation of fluxons. The impedance of the line and the resistance are selected in a way that matches the impedance of the stacked SQUIDs. Each loop of the stacked SQUIDs is magnetically coupled to one of the RL circuits coming out from SFQ-to-DC intrinsic (and smaller) SQUID loop.

The SQUID cells are in series and biased by the same DC current bias line. When an SFQ pulse is coupled to any SQUID cell, it will switch and generate a $\Phi_0$ flux quantum. Since all SQUIDs are in series, the voltage at the bias terminal is the cumulative voltage of all SQUIDs, hence the multiple of a single SQUID voltage by the number of stages.

The schematic was optimized for the input impedance of the CONNECT cell library and an output impedance of 50 Ω. After optimization, the layout of the circuit was designed for the standard process STP2 and fabricated as shown in Figure 2. This layout was designed with four stages corresponding to an amplification by a factor of four and covers an area of 160 μm×320μm.

## 3. Simulation

The schematic of the circuit was designed and the netlists were simulated using the JSIM [28] program. The optimization of the cells was done by the method of center of gravity used in the AUTO5 [29] code. In this method, the optimized parameters were the output SQUIDs, the SFQ-to-DC cell loops and the magnetic couplings between SQUIDs and SFQ-to-DC cells. Figure 3 shows the simulation output for 2-stage and 4-stage voltage multipliers loaded with an ideal load and with no thermal noise on resistors.

However, in presence of noise and when the load is not ideal, the parasitic inductances and resistances act as a low pass filter and the corresponding simulated outputs are shown in Figure 4. We see that the output voltage has a longer setup time between low and high levels, which increases with the number of stages. This limits the gain-bandwidth product. For this technology, the output levels start to interfere with each other when going over ~25 GHz for the 4-stage voltage multiplier.

The bit error rate (BER) of the 4-stage voltage multiplier was estimated from JSIM simulations for a range of current bias in presence of noise [30], as shown in Figure 5. The bias margin is calculated for the output SQUID stage only. Since all SQUIDs are in series, the margin stays the same as the number of stages is increased.

To confirm the ability to stack more stages, a 16-level output stage was designed, and the result circuit was simulated (Figure 6). The trade-off in increasing the number of stages is the decrease in output speed and increase of circuit size and bias value. The achieved gain from this circuit is 24 dB and the output frequency for this gain can reach 10 GHz. Figure 7 demonstrates the simulated gain and bandwidth of the amplifiers based on the number of stages used in the design. The minimum bandwidth, of about 10 GHz, is determined by the number of stages corresponding to the gain reaching an asymptotic maximum.

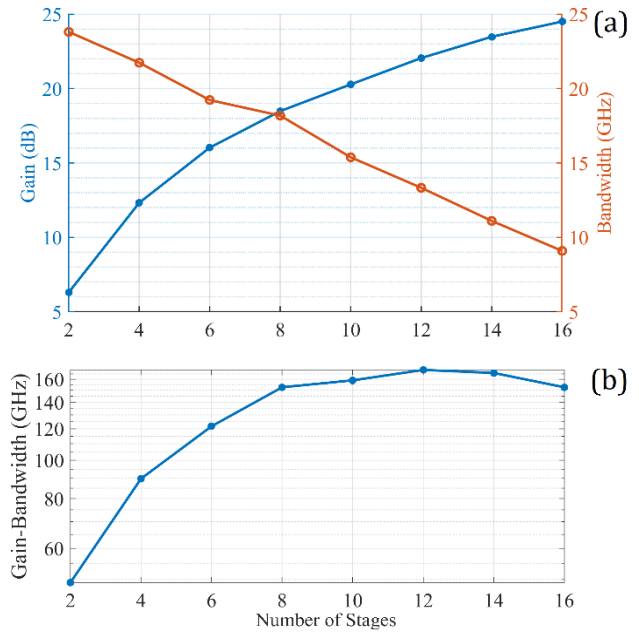

Figure 7 (a) Simulated gain and bandwidth, and (b) gain-bandwidth product, of the superconductor amplifiers, as a function of the number of stages used in the design. We observe that the gain-bandwidth product is maximum from 8 stages.

## 4. Experimental Results

The circuits were fabricated in the AIST CRAVITY Foundry, based on the 2.5kA standard niobium process (STP2 [7]) and were tested in a closed-cycle pulse-tube cooler at 4.2 K inside a shielded room [31]. Figure 8 shows the closed cycle cooler system and the wiring used for measurements.

The measurements, carried out on 2-stage and 4-stage multipliers, were done by applying a rectangular signal through a low-pass filter at low frequency to confirm the operation of the circuits. Figure 9 shows the result of the measurement at 5 kHz input signal. It should be mentioned that the input signal is fed into the DC-to-SFQ converter which is triggered at the rising edges of the input signal. Since the output changes state for each input triggering event, the output frequency is half the input signal frequency. The generated SFQ pulses are still at ~200 μV of amplitude.

By comparing the output of the amplifier for two and four stages from Figure 9 with the simulation results from Figure 4, we observed that the simulation results are in a very good agreement with the experiments.

Finally, to confirm the design with measurements, the bit error rate of the 4-stage multiplier was measured. Figure 10

shows the BER value of the circuit versus the normalized current bias value of the multiplier stage. For each point 5000 samples were recorded. Both simulation and experimental results show the same value of output voltage of 0.3 mV and 0.6 mV at two and four stages respectively and the same margin of ±2.5% for the SQUID stack bias value. The ±2.5% margin is the bias margin for the DC-SQUID stack. This current is applied separately with an external precision power supply. Therefore this margin does not limit the proper operation of the amplifier and the circuit. The digital circuit bias margin in the CONNECT library is over ±20% as mentioned in [27].

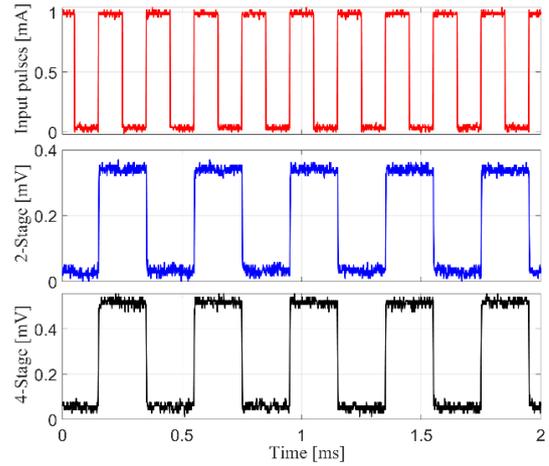

Figure 9 Experimental results for the voltage multiplier circuit. The output is differential and was measured at a frequency of 5kHz.

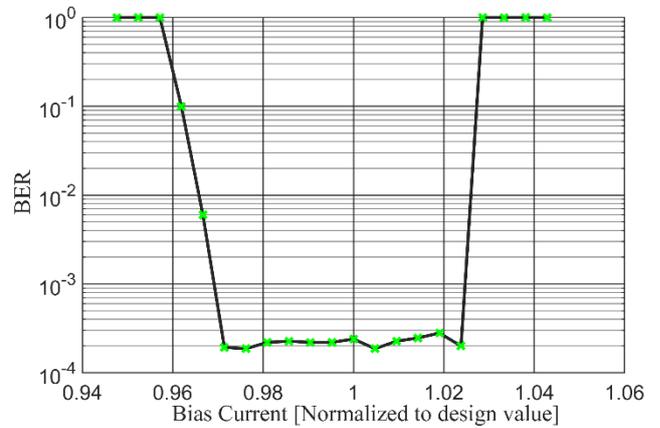

Figure 10 Bit error rate measurement of the 4-stage voltage multiplier versus normalized DC-SQUIDs bias. The error floor is the digitization error.

## 5. Conclusion

A stackable flux-multiplier circuit was designed for better interfacing between SFQ and CMOS circuits. We fabricated and tested a 4-stage stack. Experimental results are in very good agreement with simulations. The circuits were tested in a closed-cycle system to show the robustness for commercial applications. The size of the circuit can be much smaller, and margins can increase by custom cell design. To show the stackable ability of the design, a 16-stage amplifier was simulated and shows 24 dB gain up to 10 GHz. Due to quantized nature of amplification there is no added noise. This circuit increases the interface speed between SFQ and CMOS and decrease the cost of the amplification stage. The higher speed, lower noise and no dissipation of the circuit make it an ideal candidate for readout stage of quantum circuits.

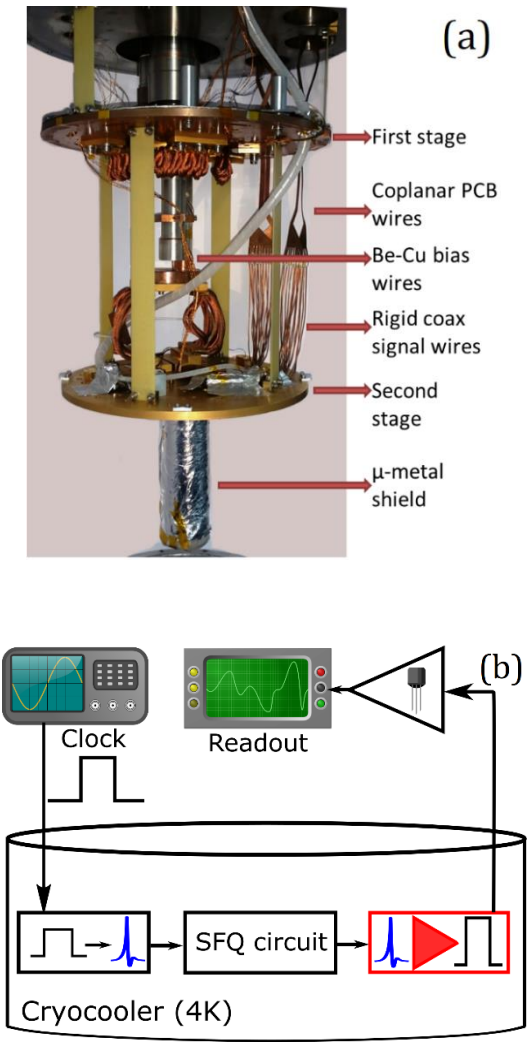

Figure 8 a) The picture of cryocooler with wirings and shield. b) Test setup for this experiment. The red part is the subject of this work.

**Acknowledgements**

This work has been partly funded by the French-Turkish Partenariat Hubert Curien (PHC) BOSPHORE No: 39708XA and TUBITAK under grant 117F266 and 117E816. The circuits are fabricated in the clean room for analog-digital superconductivity (CRAVITY) of the National Institute of Advanced Industrial Science and Technology (AIST). The research is partly based upon work supported by the Office of the Director of National Intelligence (ODNI), Intelligence Advanced Research Projects Activity (IARPA), via the U.S. Army Research Office grant W911NF-17-1-0120. The views and conclusions contained herein are those of the authors and should not be interpreted as necessarily representing the official policies or endorsements, either expressed or implied, of the ODNI, IARPA, or the U.S. Government. The U.S. Government is authorized to reproduce and distribute reprints for Governmental purposes notwithstanding any copyright notation herein.